\documentclass[aps,pra,twocolumn,superscriptaddress,showpacs,showkeys,amsmath,amssymb]{revtex4}

\usepackage{amsfonts}
\usepackage{amssymb,amsmath}
\usepackage{mathrsfs}
\usepackage{latexsym}
\usepackage{amsmath}
\usepackage[cp1251]{inputenc}
\usepackage{graphicx}
\usepackage{dcolumn}

\usepackage{bm}
\usepackage{color}
\usepackage[english]{babel}

\RequirePackage{ifthen}
\RequirePackage[pdfstartview=FitH]{hyperref}
\begin{document}

    \title{Two- and three-body effective potentials between impurities in ideal BEC}

    \author{G.~Panochko}
    \affiliation{Department of Optoelectronics and Information Technologies, Ivan Franko National University of Lviv, 107 Tarnavskyj Str., Lviv, Ukraine}
    \author{V.~Pastukhov\footnote{e-mail: volodyapastukhov@gmail.com}}
    \affiliation{Department for Theoretical Physics, Ivan Franko National University of Lviv, 12 Drahomanov Str., Lviv, Ukraine}

    \date{\today}



    \begin{abstract}
    We exactly calculate the full temperature dependence of Casimir-like forces appearing between two and three static impurities loaded in the ideal Bose gas below the Bose-Einstein condensation transition point. Assuming the short-ranged character of the boson-impurity interaction, the calculation procedure presented here can be easily extended on a Bose system with an arbitrary number of impurities immersed.
    \end{abstract}

    \maketitle

\section{Introduction}
\label{sec1}
\setcounter{equation}{0}

The presence of impurities in Bose-Einstein condensate (BEC) predetermines a number of experimentally observable phenomena, namely, formation of a single-impurity dressed quasiparticles which are referred to Bose polarons at low temperatures in one- \cite{Catani} and three-dimensional \cite{Jorgensen, Hu} systems and at finite temperatures near the BEC transition point \cite{Yan}; creation of the Rydberg polaron \cite{Camargo} in Sr condensate. A single ion immersed in BEC may be also used to probe \cite{Schmid} the local atomic density distribution of host atoms. Theoretical investigations of properties of such objects at finite temperatures are mostly focused on the exploration of the Bose polaron behavior near the superfluid phase-transition point \cite{Levinsen1,Guenther,Pastukhov2018}, and on the investigation of impurity dynamics \cite{Boudjemaa,Liu}. Polarons can be also used for the low-temperature thermometry in BECs \cite{Mehboudi}.

In real experimental conditions, however, the number of impurities is macroscopic nevertheless concentrations are typically small. Therefore, even when impurities are initially non-interacting (spin-polarised fermions, for instance), being immersed in the Bose environment they mutually interact via the effective boson-mediated potential. In general, this effective potential is the many-body one but when the concentration of impurities is small (i.e., average distance between exterior particles is large) it can be freely modelled by the pairwise interactions. Depending on a strength of the boson-impurity coupling the two-impurity system undergoes crossover behavior from two separate Bose polarons interacting via weak Yukawa-like potential \cite{Naidon} through the bipolaron state \cite{Tempere,Ardila} at intermediate couplings to the Efimov trimer \cite{Naidon,Levinsen2,Sun} at strong boson-impurity attractions. It is interesting that the binding energy of trimers at unitary is suppressed \cite{Zinner2013,Zinner2014} by presence of a Bose condensate.

The medium-induced Casimir-like forces are of great importance in the condensed-matter physics \cite{Klimchitskaya}. Being responsible for our understanding of numerous phenomena in many-body systems they have the most profound effect in low dimensions, particularly in 1D \cite{Recati,Wachter,Kamenev,Rodin}. In this context the simplest system for the visualization of the Casimir forces is two particles interacting with a free scalar field. This is a model of 1D crystal in the harmonic approximation, where the scattering of phonons on impurities leads to the induced long-range interaction between them. A behavior of the effective potential essentially differs for impurities with finite and infinite masses \cite{Pavlov2018}, namely, $\Phi_{\rm eff}(R_{12})\propto 1/R^3_{12}$ and $\Phi_{\rm eff}(R_{12})\propto 1/R_{12}$, respectively, and becomes exponential at finite temperatures \cite{Pavlov2019}. The appearance of interaction of the Casimir type can be also demonstrated in mixtures of quantum gases \cite{Salvo}, where the effective attraction between `heavy' bosons of $^{133}$Cs mediated by the degenerated Fermi gas of $^6$Li atoms was observed. Theoretically this problem was studied in Ref.~\cite{Nishida} for both 2D and 3D cases.

The aim of the present article is to explore the finite-temperature Casimir effect associated with the immersion of impurities in a 3D ideal Bose gas. In general, a problem of induced forces in Bose systems is not well-studied, especially at finite temperatures. Few exceptions are the following: the perturbative consideration of the effective interaction between static impurities in the spin-orbit-coupled BEC \cite{Song}, the detailed discussion of the Landau effective potential for two Bose polarons at absolute zero \cite{Guardian}, and systems confined in 1D (or quasi-1D) geometries, which are now lively discussed and where the peculiarities of the induced interaction in the dilute limit are dictated by the characteristic scale, namely, the coherence length. For distances between impurities less than this scale the Casimir force behaviors exponentially \cite{Dehkharghani}, while decaying power-law-like at large inter-particle spacing \cite{Reichert1} with the boundary-conditions-dependent exponent \cite{Reichert3}. At that time, the finite-temperature fluctuations not only break the quasi-long-range order in 1D bosonic systems but also change the large-distance behavior of the Casimir force to exponential \cite{Reichert2}.

\section{Formulation}
\subsection{Statement of the problem}
We consider a very simple model of a few static (infinite-mass) impurities immersed in ideal BEC. Although below we mainly focus on the one-, two- and three-particle limits the general calculation scheme is also applicable for an arbitrary number of impurities. Therefore, the Hamiltonian of the system 
\begin{eqnarray}\label{H}
H=\sum_{{\bf k}, {\bf q}} \langle {\bf k}|\varepsilon+\Phi({\bf r})|{\bf q}\rangle\psi^+_{\bf k}\psi_{\bf q},
\end{eqnarray}
is written down for this latter case, where $\varepsilon=\frac{{\bf p}^2}{2m}$ is the one-boson kinetic energy operator, and
\begin{eqnarray}\label{Phi}
\Phi({\bf r})=\sum_{1\le j\le \mathcal{N}}g\delta({\bf r}-{\bf R}_j),
\end{eqnarray}
is the potential energy due to interaction with $\mathcal{N}$ impurities placed in positions ${\bf R}_j$. Operators $\psi^+_{\bf k}$ ($\psi_{\bf k}$) are standard creation (annihilation) bosonic operators of particle with momentum $\hbar {\bf k}$, and in (\ref{H}) we used the plane-wave representation defined for large volume $V$ with periodic boundary conditions imposed. The strength of boson-impurity interaction is controlled by bare coupling constant $g$, which because of $\delta$-type interaction should be renormalized in final formulas $\frac{1}{g}=\frac{m}{2\pi\hbar^2a}-\frac{1}{V}\sum_{{\bf k}}\frac{1}{\varepsilon_k}$ via $s$-wave scattering length $a$.

The thermodynamics of non-interacting bosons loaded in external potential (\ref{Phi}) can be obtained in conventional \cite{Landau_5} for an ideal quantum gases way. First, we have to solve the single-particle quantum mechanical problem
\begin{eqnarray}\label{h}
\{\varepsilon+\Phi({\bf r})\}|l)=\mathcal{E}_l|l),
\end{eqnarray}
(here $|l)$ and $\mathcal{E}_l$ are the $l$-th eigenstate and corresponding eigenvalue of a single-boson Hamiltonian in the presence of impurities) and then straightforwardly apply the grand-canonical formalism with grand potential given by
\begin{eqnarray}\label{Omega}
\Omega=T\sum_l \ln\left[1-e^{(\mu-\mathcal{E}_l)/T}\right],
\end{eqnarray}
where $\mu$ is the chemical potential that controls the number of bosons. The sum in r.h.s. of Eq.~(\ref{Omega}) is a trace of the one-particle statistical operator and taking into account the invariance of trace we can equivalently rewrite $\Omega$ in the plane-wave basis [from now on we do not write down in $\Phi({\bf r})$ the explicit dependence on {\bf r}]
\begin{eqnarray*}
\Omega=T\sum_{{\bf k}} \langle {\bf k}|\ln\left[1-e^{(\mu-\varepsilon-\Phi)/T}\right]|{\bf k}\rangle,
\end{eqnarray*}
or in a form more convenient for practical use
\begin{eqnarray}\label{Omega_omega}
	\Omega=T\int {d\omega} D(\omega)\ln\left[1-e^{(\mu-\omega)/T}\right],
\end{eqnarray}
where the density of states $D(\omega)=\sum_l \delta(\omega-\mathcal{E}_l)=\sum_{{\bf k}}\langle{\bf k}|\delta(\omega -\varepsilon-\Phi)|{\bf k}\rangle$ can be conventionally written through the one-particle Green's function $G_{\omega}$
\begin{eqnarray}\label{D_omega}
D(\omega)=-\frac{1}{\pi}\sum_{{\bf k}}  {\rm Im}\langle{\bf k}|G_{\omega+i0}|{\bf k}\rangle, \ \  G_{\omega}=\frac{1}{\omega-\varepsilon-\Phi},
\end{eqnarray}
and integration is carried out in the semi axis where $\omega-\mu>0$. So, the further consideration is fully devoted to calculations of the above-presented density of states. Furthermore, in the following we restrict ourselves to the thermodynamic limit, where both volume $V$ and number of bosons $N$ rapidly grow, while keeping density $n=N/V$ of the system fixed. Demanding additivity of the thermodynamic potential we assume that impurity does not change properties of Bose gas drastically, which particularly means that only scattering states will be accounted. The appearance of bound states at some region of parameters change will immediately lead to the collapse of bosons (in that case the macroscopic number of particles will be localized in a finite volume).

All peculiarities of the density of states can be figured out by computing the $T$-matrix
\begin{eqnarray}\label{Tau}
\mathcal{T}_{\omega}=\Phi+\Phi G^{(0)}_{\omega}\mathcal{T}_{\omega},
\end{eqnarray}
which allows to represent Green's function in terms of its zero-order counterpart $ G^{(0)}_{\omega}$
\begin{eqnarray}\label{G}
G_{\omega}= G^{(0)}_{\omega}+ G^{(0)}_{\omega}\mathcal{T}_{\omega} G^{(0)}_{\omega}, \ \  G^{(0)}_{\omega}=\frac{1}{\omega-\varepsilon}.
\end{eqnarray}
Working with BECs and assuming the scattering nature of the ground state in a Bose gas with impurities, the zero-momentum term in formula for $D(\omega)$ should be treated with a great care. First, it is more instructive to rewrite it as follows \cite{Luscher}
\begin{eqnarray}\label{0_G_0}
\langle{\bf 0}|G_{\omega+i0}|{\bf 0}\rangle=\frac{1}{\omega+i0-\langle {\bf 0}|\mathcal{T}^{\bf 0}_{\omega+i0}|{\bf 0}\rangle},
\end{eqnarray}
where the reduced $T$-matrix $\mathcal{T}^{\bf 0}_{\omega}=\Phi+\Phi G^{(0)}_{\omega}Q_{\bf 0}\mathcal{T}^{\bf 0}_{\omega}$, (here $Q_{\bf 0}=1-|{\bf 0}\rangle\langle {\bf 0}|$ denotes the projector on all scattering states without ${\bf k}={\bf 0}$) is determined on the subspace of the one-particle Hilbert space without state $|{\bf 0}\rangle$. Substitution in Eq.~(\ref{Omega_omega}) leads us to the conclusion that the chemical potential of bosons with microscopic number of impurities reads
\begin{eqnarray}\label{mu}
\mu=\langle {\bf 0}|\mathcal{T}^{\bf 0}_{0}|{\bf 0}\rangle.
\end{eqnarray}
Note that $\mu$ is of order $1/V$ when only a few particles are immersed in a system below the critical temperature, and Eq.~(\ref{mu}) guarantees only a leading-order asymptotics in $1/V$. Multiplied by the number of bosons $N$, the result (\ref{mu}) represent the impurities binding energy at $T=0$. So, the problem is reduced to the calculations of the $T$-matrix of a bosonic atom moving in the external potential produced by the static particles. Fortunately, the formal solution of this problem can be exactly found for arbitrary non-macroscopic number of motionless impurities. Particularly, by rewriting the operator equation (\ref{Tau}) (with projector $Q_{\bf 0}$ inserted) explicitly in matrix form we obtain
\begin{align}\label{Tau_qk}
	&\langle {\bf q}|\mathcal{T}^{\bf 0}_{\omega}|{\bf k}\rangle=\frac{1}{V}\sum_{1\le j\le \mathcal{N}} g e^{-i({\bf q}-{\bf k})\cdot{\bf R}_j}\nonumber\\
	&+\frac{1}{V}\sum_{{\bf s}\neq 0}\sum_{1\le j\le \mathcal{N}} g e^{-i({\bf q}-{\bf s})\cdot{\bf R}_j}\frac{1}{\omega-\varepsilon_s}\langle {\bf s}|\mathcal{T}^{\bf 0}_{\omega}|{\bf k}\rangle.
\end{align}
Applying the iterative procedure to the above equation, it is easy to guess the solution for the $T$-matrix $\langle {\bf q}|\mathcal{T}^{\bf 0}_{\omega}|{\bf k}\rangle=t_{\omega}\frac{1}{V}\sum_{1\le i,j\le \mathcal{N}}e^{-i{\bf q}\cdot{\bf R}_i}T_{ij}e^{i{\bf k}\cdot{\bf R}_j}$, where the elements of quadratic matrix $T_{ij}$ (of size $\mathcal{N}$) are defined as follows
\begin{eqnarray}\label{T_ij}
T^{-1}_{ij}=\delta_{ij}-\Delta_{\omega}(R_{ij})[1-\delta_{ij}].
\end{eqnarray}
Here $\delta_{ij}$ is the Kronecker delta and we made use of notations
\begin{eqnarray}\label{t}
t^{-1}_{\omega}=\frac{m}{2\pi\hbar^2a}-\frac{1}{V}\sum_{{\bf k}\neq 0}\left(\frac{1}{\omega-\varepsilon_k}+\frac{1}{\varepsilon_k}\right),
\end{eqnarray}
for the boson-single-impurity vacuum $T$-matrix and for function 
\begin{eqnarray}\label{Delta_R}
\Delta_{\omega}(R)=t_{\omega}\frac{1}{V}\sum_{{\bf k}\neq 0}\frac{e^{i{\bf k}\cdot{\bf R}}}{\omega-\varepsilon_k},
\end{eqnarray} 
of relative distance between impurities ${\bf R}_{ij}={\bf R}_i-{\bf R}_j$.

In general for fixed $\mathcal{N}$, the problem is reduced to cumbersome matrix calculus, but in the two-impurity limit these calculations can be carried out comparatively simply,
\begin{eqnarray}\label{Tau_2}
V\langle {\bf k}|\mathcal{T}^{{\bf 0}}_{\omega}|{\bf k}\rangle=2t_{\omega}\frac{1+\Delta_{\omega}(R_{12})\cos({\bf k}\cdot{\bf R}_{12})}{1-\Delta^2_{\omega}(R_{12})}.
\end{eqnarray}
The evaluation of the above integrals in three spacial dimensions causes any problems and can be found in Appendix. With Eq.~(\ref{Tau_2}) in hands, which actually gives the density of states, we are free to calculate thermodynamics of the considered system. The solution of Eq.~(\ref{Tau_qk}) for three exterior particles immersed reads
\begin{align}\label{V_kk}
& V\langle {\bf k}|\mathcal{T}^{{\bf 0}}_{\omega}|{\bf k}\rangle=\frac{t_{\omega}}{\textrm{det}_{\omega}(R_{12},R_{13},R_{23})}\nonumber\\ 
&\times\Bigg\{3-\Delta^{2}_{\omega}({R}_{12})
-\Delta^{2}_{\omega}({R}_{13})-\Delta^{2}_{\omega}({R}_{23})\nonumber\\
&+2[\Delta_{\omega}({R}_{12})+\Delta_{\omega}({R}_{13})\Delta_{\omega}({R}_{23})]\cos{(\bf{k}\cdot{\bf{ R}}_{12})}\nonumber\\
&+2[\Delta_{\omega}({R}_{13})+\Delta_{\omega}({R}_{12})\Delta_{\omega}({R}_{23})]\cos{(\bf{k}\cdot{\bf{ R}}_{13})}\nonumber\\
&+2[\Delta_{\omega}({R}_{23})+\Delta_{\omega}({R}_{12})\Delta_{\omega}({R}_{13})]\cos(\bf{k}\cdot{\bf R}_{23})\Bigg\},
\end{align}
here $\det_{\omega}(R_{12},R_{13},R_{23})$ is the determinant of matrix $T^{-1}_{ij}$
\begin{eqnarray}\label{det_omega}
&&\textrm{det}_{\omega}(R_{12},R_{13},R_{23})=1-\Delta^{2}_{\omega}({R}_{12})-\Delta^{2}_{\omega}({R}_{23})\nonumber\\
&&-\Delta^{2}_{\omega}({R}_{13})-2\Delta_{\omega}({R}_{12})\Delta_{\omega}({R}_{23})\Delta_{\omega}({R}_{13}).
\end{eqnarray}

The ground-state energy of our system can be also calculated in the plane-wave basis by applying the conventional many-body perturbation techniques directly to Hamiltonian (\ref{H}). With the assumption that presence of impurities does not destroy uniformity of the Bose condensate and the lowest one-particle energy level again corresponds to wave-vector ${\bf k}={\bf 0}$, the Hamiltonian reads
\begin{eqnarray}\label{H_cond}
H=N\langle {\bf 0}|\Phi|{\bf 0}\rangle+\sum_{{\bf k}, {\bf q}\neq 0} \langle {\bf k}|\varepsilon+\Phi|{\bf q}\rangle\psi^+_{\bf k}\psi_{\bf q}\nonumber\\
+\sqrt{N}\sum_{{\bf k}\neq 0} \left\{\langle {\bf k}|\Phi|{\bf 0}\rangle\psi^+_{\bf k}+\langle {\bf 0}|\Phi|{\bf k}\rangle\psi_{\bf k}\right\},
\end{eqnarray}
where both $\psi^+_{\bf 0}$, $\psi_{\bf 0}$ are replaced by a $c$-number $\sqrt{N}$.
Calculated to all orders of perturbation theory, the ground-state energy
\begin{align}\label{E_0}
&E_0=N\langle {\bf 0}|\Phi|{\bf 0}\rangle+N\sum_{{\bf k}\neq 0}\langle {\bf 0}|\Phi|{\bf k}\rangle \frac{1}{-\varepsilon_k}\langle {\bf k}|\Phi|{\bf 0}\rangle\nonumber\\
&+N\sum_{{\bf k}, {\bf q}\neq 0} \langle {\bf 0}|\Phi|{\bf k}\rangle \frac{1}{-\varepsilon_k}\langle {\bf k}|\Phi|{\bf q}\rangle\frac{1}{-\varepsilon_q}\langle {\bf q}|\Phi|{\bf 0}\rangle+\ldots\nonumber\\
&=N\langle {\bf 0}|\mathcal{T}^{\bf 0}_{0}|{\bf 0}\rangle,
\end{align}
collapses exactly to the diagonal element of the reduced $T$-matrix and reproduces (\ref{mu}).

The internal energy of the system and the average number of bosons, at temperatures above the BEC transition, are derived by applying thermodynamic relations $E=\Omega+\mu N+TS$ [with $S=-\left(\partial \Omega/\partial T\right)_{V,\mu}$ being the entropy of the system] and $N=-\left(\partial \Omega/\partial \mu\right)_{V,T}$, respectively
\begin{eqnarray}\label{E_and_N}
E=\int \frac{{d\omega} D(\omega)\omega}{e^{(\omega-\mu)/T}-1}, \ \ N=\int \frac{{d\omega}D(\omega)}{e^{(\omega-\mu)/T}-1},
\end{eqnarray}
where $\mu$ contains both (\ref{mu}) and the finite temperature-dependent part. Below the transition temperature, the calculations of integrals in (\ref{E_and_N}) require separation of the ${\bf k}={\bf 0}$-mode in $D(\omega)$. Taking into account Eqs.~(\ref{0_G_0}), (\ref{mu}) and the second formula in (\ref{E_and_N}) we conclude that $D(\omega)$ contains $\delta$-singularity at $\omega=\mu$ (i.e., at $\omega\propto 1/V$), such that
\begin{eqnarray}
N=N_0+\int \frac{{d\omega}D^{\bf 0}(\omega)}{e^{(\omega-\mu)/T}-1},
\end{eqnarray}
where $N_0$ is the number of particles in BEC and superscript near $D(\omega)$ denotes that term with ${\bf k}={\bf 0}$ omitted in the density of states. Same substitution of $D(\omega)$ in equation for the internal energy below the BEC transition then yields
\begin{eqnarray}\label{E}
E=N_0\langle {\bf 0}|\mathcal{T}^{\bf 0}_{0}|{\bf 0}\rangle+\int \frac{{d\omega} D^{\bf 0}(\omega)\omega}{e^{(\omega-\mu)/T}-1}.
\end{eqnarray}
The latter expression can be equivalently obtained by means of the quasi-particle-picture arguments. Indeed, the first term in (\ref{E}) represents the contribution of BEC (where each boson now has an energy $\mu$ due to presence of impurities) to the total energy, while the second term in $E$ is the average energy of the thermally-excited Bose particles. Equation (\ref{E}) allows to obtaining (see Appendix) the exact energy that Bose gas gains when $\mathcal{N}$ impurities are immersed in it at finite temperatures.

\subsection{Bound states}
In order to elucidate the limits of applicability of the above formal calculations 
we must analyze the one-boson bound-state problem. This can be directly done by searching for the $T$-matrix (\ref{Tau_2}) poles at negative $\omega$s. For a single impurity, they are given by zeros of $t^{-1}_{\omega}$, which lead to the fictitious pole with a simple mathematical expression $\epsilon_{1}=-\frac{\hbar^2}{2ma^2}$, valid for all positive $a$s. The appropriate one-boson Hamiltonian and the ground-state function read
\begin{eqnarray}\label{H_1}
H&=&-\frac{\hbar^2}{2m}\nabla^2+\frac{2\pi\hbar^2a}{m}\delta({\bf r}_1)\frac{\partial}{\partial r_1}r_1, \\
&& \langle {\bf r}|0)_{\mathcal{N}=1}\propto \frac{e^{-r_1/a}}{{r_1}},
\end{eqnarray}
with shorthand notation ${\bf r}_1={\bf r}-{\bf R}_1$ for relative boson-impurity position. The above potential energy is the well-known Huang-Yang pseudo-potential, which explicitly represents the Bethe-Peierls boundary condition. In the two-impurity case, the situation with the bound states is more interesting. Now poles, $\epsilon_2$, corresponding to bound states are given by two equations \cite{Zinner2013}
\begin{eqnarray}\label{epsilon_2}
1-a\sqrt{2m|\epsilon_2|}/\hbar\pm \frac{a}{R_{12}}e^{-R_{12}\sqrt{2m|\epsilon_2|}/\hbar}=0.
\end{eqnarray}
For completeness, we also provide the bound-state wave functions \cite{Nishida} [the appropriated Hamiltonian in this case is just a two-centered generalization of (\ref{H_1})]
\begin{eqnarray}\label{2_wave_func}
\left.\begin{array}{c}
\langle {\bf r}|0)_{\mathcal{N}=2}\\
\langle {\bf r}|1)_{\mathcal{N}=2}
\end{array}\right\}\propto \frac{e^{-r_1\sqrt{2m|\epsilon_2|}/\hbar}}{{r_1}}\pm \frac{e^{-r_2\sqrt{2m|\epsilon_2|}/\hbar}}{{r_2}}.
\end{eqnarray}
A graphical representation of solutions for dimensionless quantity $|\tilde{\epsilon}_2|=|\epsilon_2|/\left(\frac{\hbar^2}{2mR^2_{12}}\right)$ is plotted in Fig.~1 
\begin{figure}[h!]
	\includegraphics[width=0.5\textwidth,clip,angle=-0]{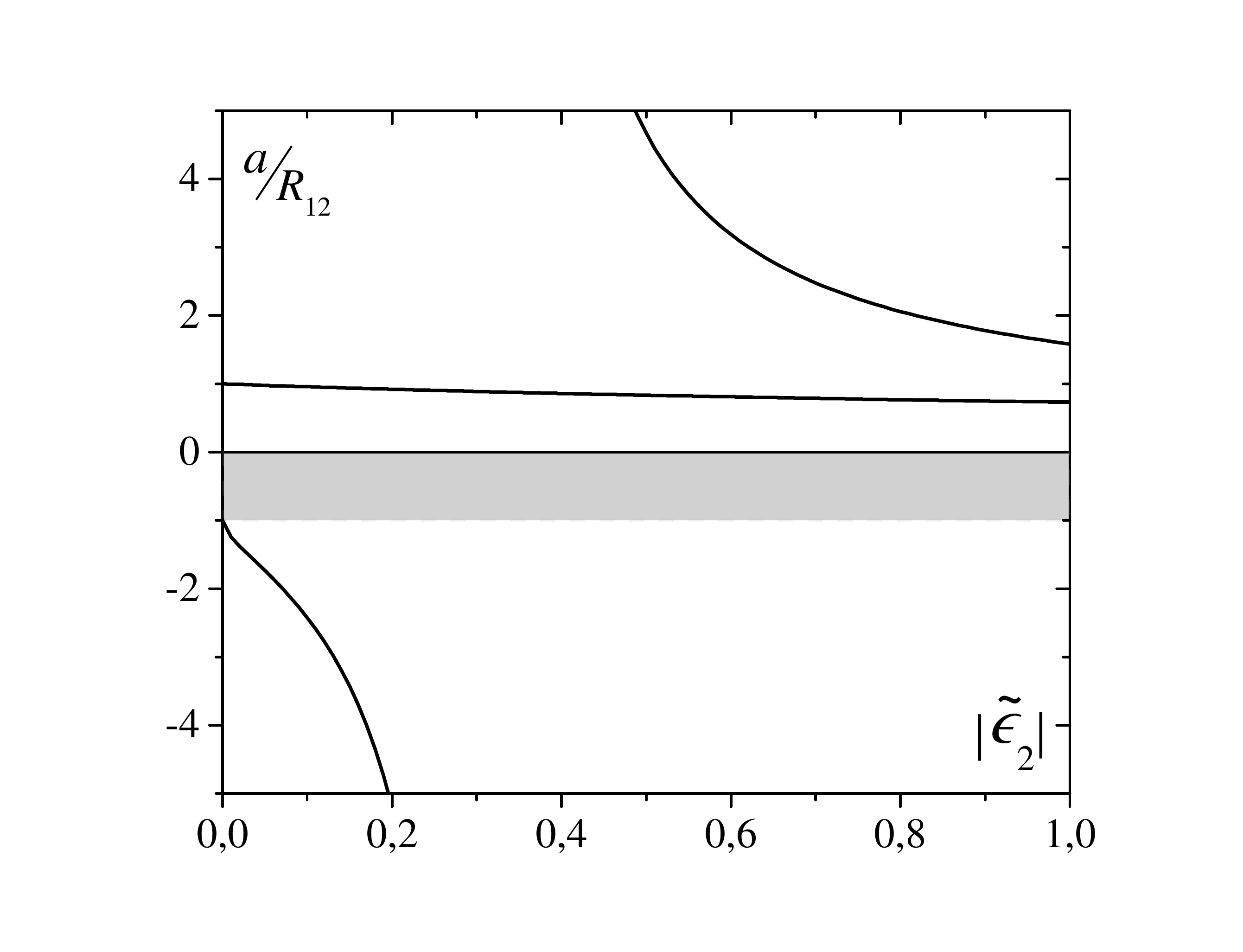}
	\caption{Graphical representation of Eqs.~(\ref{epsilon_2}) for the bound-state energies of bosons with two impurities immersed. At positive $a$ and large $R_{12}>a$ there are two energy levels. In all other cases, except region $-1<\frac{a}{R_{12}}\le 0$ (gray area), a single bound state exists.}
\end{figure}
from which we clearly see that there is a `window', $-1<\frac{a}{R_{12}}\le 0$, where no bound states occur and the previous analysis is valid.

A similar analysis can be performed for three heavy atoms immersed in a system of non-interacting bosons. In this case the equation for bound states of a single boson both for positive and negative scattering lengths $a$ is the following:
 \begin{eqnarray}\label{bound_3}
 &{(1-a\sqrt{2m|\epsilon_3|}/\hbar)}^3-(1-a\sqrt{2m|\epsilon_3|}/\hbar)\nonumber\\
 &\times\Big[ \left(\frac{a}{R_{12}}\right)^{2}e^{-2R_{12}\sqrt{2m|\epsilon_3|}/\hbar}+\left(\frac{a}{R_{13}}\right)^{2}e^{-2R_{13}\sqrt{2m|\epsilon_3|}/\hbar}\nonumber\\
 &+\left(\frac{a}{R_{23}}\right)^{2}e^{-2R_{23}\sqrt{2m|\epsilon_3|}/\hbar}\Big]\nonumber\\
 &+2\frac{a^3}{R_{12}R_{13}R_{23}}{e^{-(R_{12}+R_{13}+R_{23})\sqrt{2m|\epsilon_3|}/\hbar}}=0.
\end{eqnarray}
The appropriate wave functions can be obtained by tracking back formula (\ref{2_wave_func}). In general, energy ${\epsilon}_3$ depends on three dimensionless parameters; let say three relative distances between impurities $R_{12}$, $R_{13}$, $R_{23}$ in units of $|a|$ (moreover the triangle inequalities should be preserved $R_{12}+R_{13}\leq R_{23}, R_{12}+R_{23}\leq R_{13}, R_{23}+R_{13}\leq R_{12}$).
Note that Eq.~$(\ref{bound_3})$ at least contains three solutions. However, in the limiting case when the impurities are located equidistantly from each other, $R_{12}=R_{13}=R_{23}=R$, Eq.~(\ref{bound_3}) for the bound states of boson splits into two
 \begin{eqnarray}\label{bound_3_r}
& 1-a\sqrt{2m|\epsilon_3|}/\hbar+2\frac{a}{R}e^{-R\sqrt{2m|\epsilon_3|}/\hbar}=0\\
& 1-a\sqrt{2m|\epsilon_3|}/\hbar-\frac{a}{R}e^{-R\sqrt{2m|\epsilon_3|}/\hbar}=0,
\end{eqnarray}
and branch coming from the first one, $\ref{bound_3_r}$, is two times degenerated. 
It is easily seen that Eq.~(\ref{bound_3}) transforms into Eq.~(\ref{epsilon_2}), if any of $R_{ij}$ tends to infinity. The graphical dependence (Fig.~2) illustrates the behavior of dimensionless bound state energy  $|\tilde{\epsilon}_3|=|\epsilon_3|/\left(\frac{\hbar^2}{2mR^2}\right)$ of a single boson in the presence of three impurities when $R_{12}=R_{13}=R_{23}=R$. Here we see that in a case of equidistant impurities immersed in the ideal Bose gas, there is a region (shaded area in Fig.~2), where the system remains stable and its ground state is not localized. However, the shaded region of the Bose-gas stability in the three-impurity case is somewhat more narrow than in the two-impurity one (compare Fig.~1 and Fig.~2).
\begin{figure}[h!]
	\includegraphics[width=0.4\textwidth,clip,angle=-0]{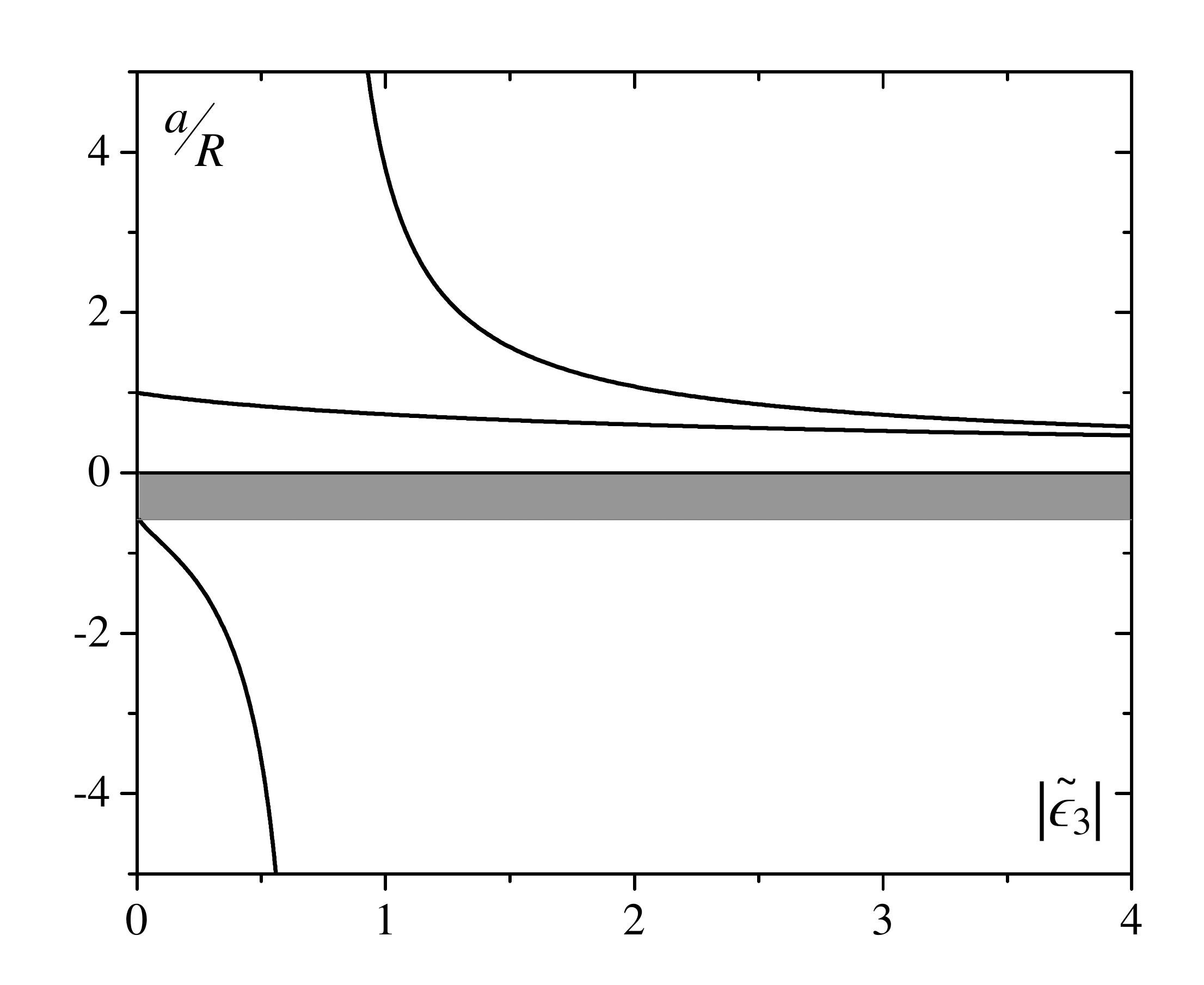}
	\caption{Graphical representation of Eqs.~(\ref{bound_3_r}) for the bound-state energies of a single boson with three equidistant impurities immersed. At positive $a$ and $R>a$ there are two (different) energy levels. In all other cases, except region $-0.5849<\frac{a}{R}\le 0$ (grey area), a single bound state exists.}
\end{figure}
If the positions of two impurities are fixed and the location of the third particle is varied we can obtain the level lines of bound state energy ${\epsilon}_3$ (see Fig.~3).
\begin{figure}[h!]
	\includegraphics[width=0.45\textwidth,clip,angle=-0]{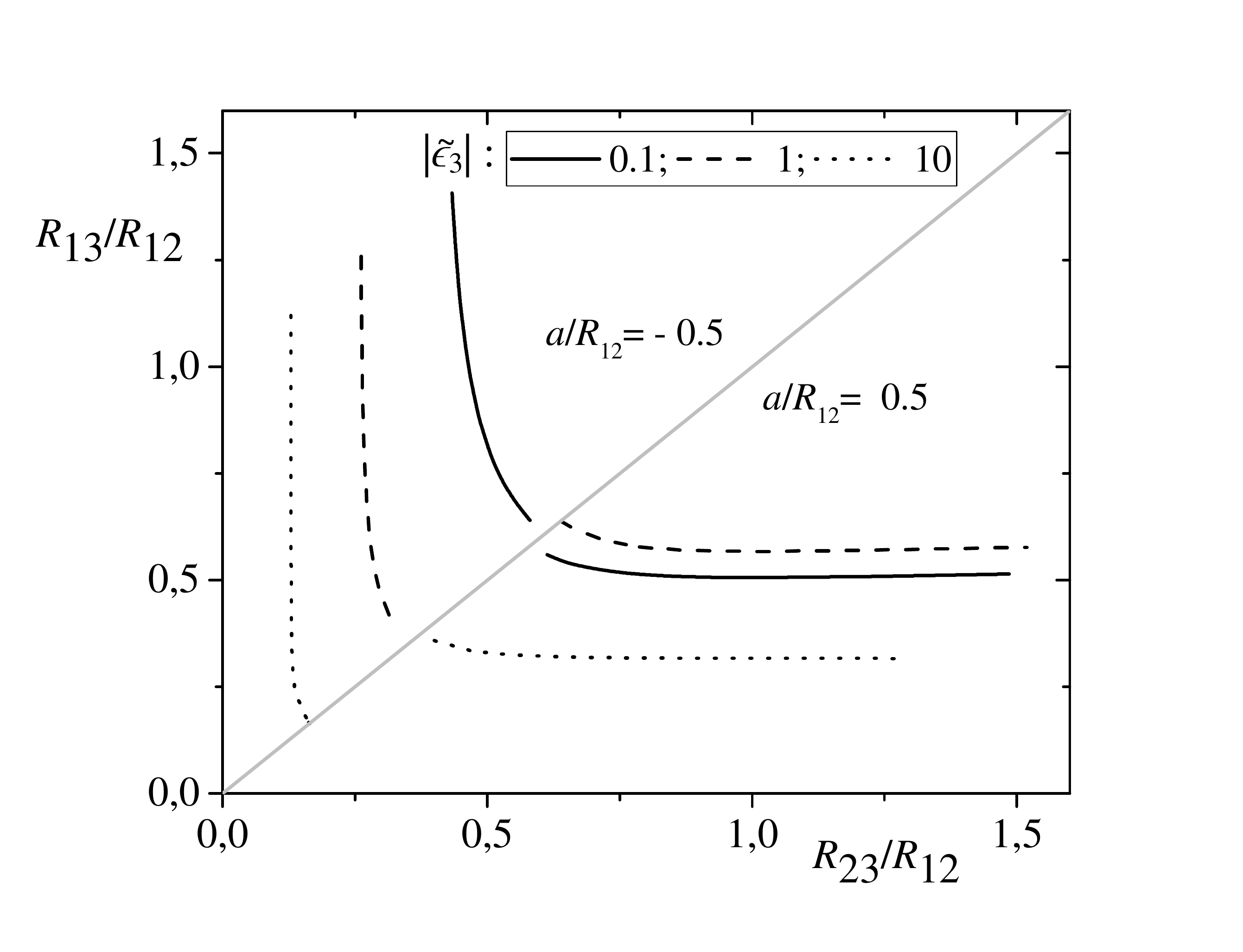}
	\caption{Graphical representation of Eqs.~(\ref{bound_3}) for the bound-state-energy level lines of a boson with three impurities immersed. The upper and lower half-planes correspond to $a/R_{12}=-0.5$ and $a/R_{12}=0.5$, respectively.}
\end{figure} 
Because of the triangle inequalities these lines are always of a finite length.
Note that in a case of negative scattering lengths ($a<0$, region above the grey line in Fig.~3) the boson bound-state energy increases with decreasing of a distances between impurities, while for $a>0$ (region below the grey line) there is a possible configuration of impurities when the level line of $|{\epsilon}_3|=0.1$ is located between the level lines of $|{\epsilon}_3|=1$ and $|{\epsilon}_3|=10$, respectively.

In the context of experimental realization at least two questions arise at this point. The first one is related to number of impurities that is typically produced in experiments and this number is significantly larger than two or three. From the previous analysis is easy to figure out that for an arbitrary number $\mathcal{N}$ of static particles the $\mathcal{N}$ wave functions that correspond to bound states are simply given by linear combinations of exponential functions of type $e^{-r_j\sqrt{2m|\epsilon_{\mathcal{N}}|}/\hbar}/r_j$. In general, these $\mathcal{N}$ energies are complicated functions of relative distances $R_{j j'}$ between impurities, but they disappear when $a<0$ and all $R_{j j'}\gg |a|$. So, whole our previous discussion is plausible for systems with small concentration of uniformly-distributed impurities. The second question raises issue of the experimental visibility of bound states. At finite temperatures the Bose gas collapse dynamics is complicated and requires separate investigation, but if we assume that the system is initially prepared at very low temperatures without boson-impurity interaction and then this interaction is suddenly switched on, probability of the bound-state realization given by modulus squared of the wave-function overlap
\begin{eqnarray}\label{Z}
Z=\left|\langle {\rm vac}|\frac{(\psi_{l=0})^N}{\sqrt{N!}}\frac{(\psi^+_{{\bf k}={\bf 0}})^N}{\sqrt{N!}}|{\rm vac}\rangle\right|^2\propto \left(\frac{a^3}{V}\right)^N,
\end{eqnarray}
is very small. Here, we have argued that system remains in the uniform (scattering) BEC state for some time even for set of parameters when true ground state is the collapsed BEC. The obtained power-law behavior of overlap (\ref{Z}) which tends to zero very quickly with increasing number of surrounding particles is usually referred to the orthogonality catastrophe. For the bosonic environments formed by non-interacting particles such a behavior seems to be generic for all spacial dimensionalities where the BEC transition occurs. The low-dimensional (already starting from two-dimensional) ideal Bose gas, instead, is insensitive to the presence of impurities.

\section{Results}
Full information about the temperature dependence of energy of two impurities can be deduced by subtracting the internal energy of ideal Bose gas from Eq.~(\ref{E}) (see Appendix). The resulting energy $\Delta E_2(R_{12})$ associated with impurities is a complicated function of relative distance $R_{12}$ but when the latter goes to infinity energy $\Delta E_2(\infty)$ tends to constant, which is twice the binding energy of a single impurity
\begin{eqnarray}\label{DeltaE1}
\Delta E_{1}=\frac{2\pi\hbar^2 a}{m}\left[n+\int^{\infty}_0 \frac{dk}{(2\pi)^2}\frac{k^2}{e^{\varepsilon_k/T}-1}\frac{(ak)^2}{1+(ak)^2}\right].
\end{eqnarray}
Typical temperature behavior of $\Delta E_{1}$ is presented in Fig.~2,
\begin{figure}[h!]
	\includegraphics[width=0.45\textwidth,clip,angle=-0]{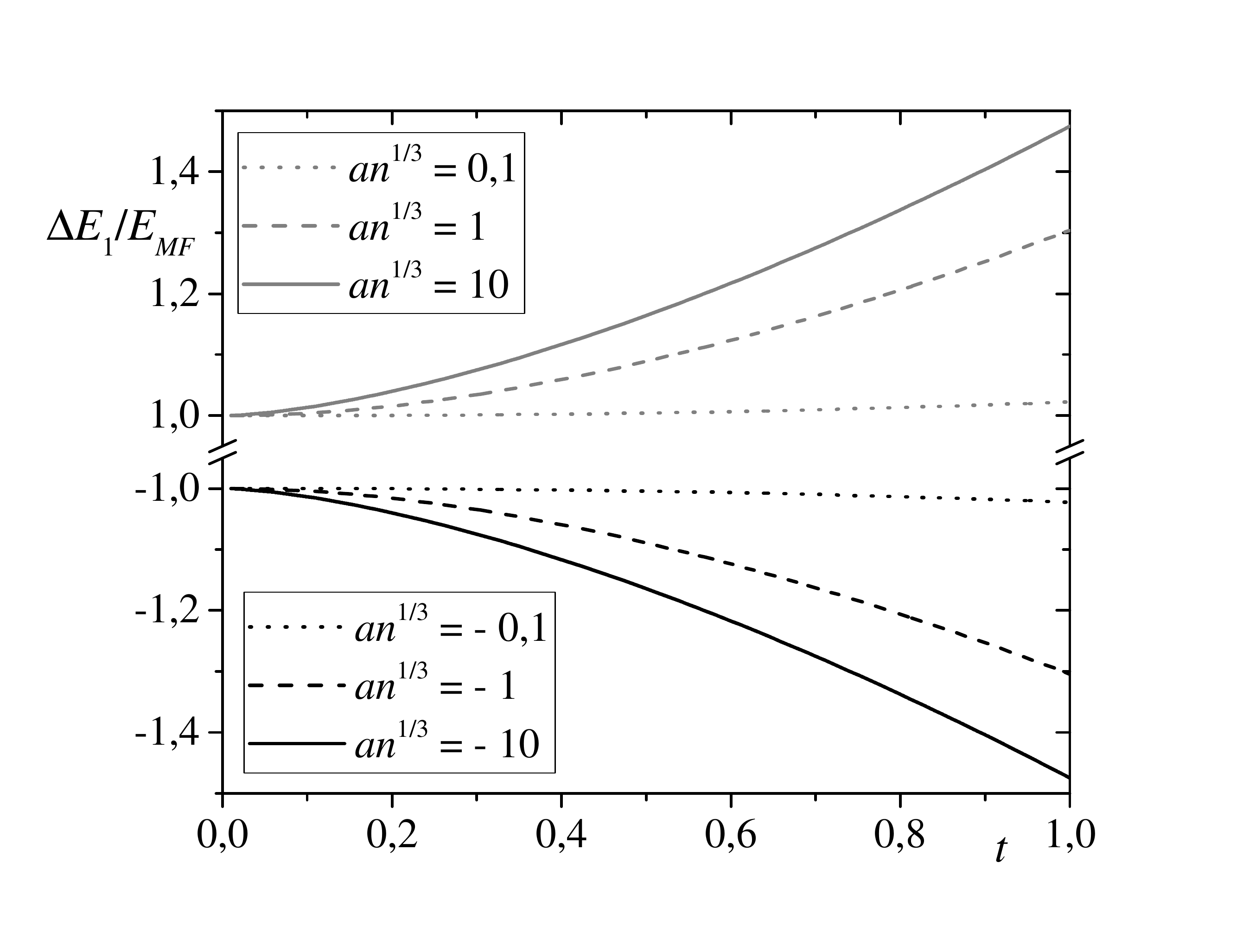}
	\caption{Examples of temperature dependence $t=T/T_0$ of the single-impurity binding energy.}
\end{figure}
where $T_0$ is the BEC temperature and $E_{{\rm MF}}=2\pi\hbar^2|a|n/m$ is a modulus of the mean-field energy (which is the exact for one impurity at $T=0$). We also built in Fig.~3 the one-particle energy at several fixed temperatures as a function of dimensionless coupling constant $an^{1/3}$.
\begin{figure}[h!]
	\includegraphics[width=0.45\textwidth,clip,angle=-0]{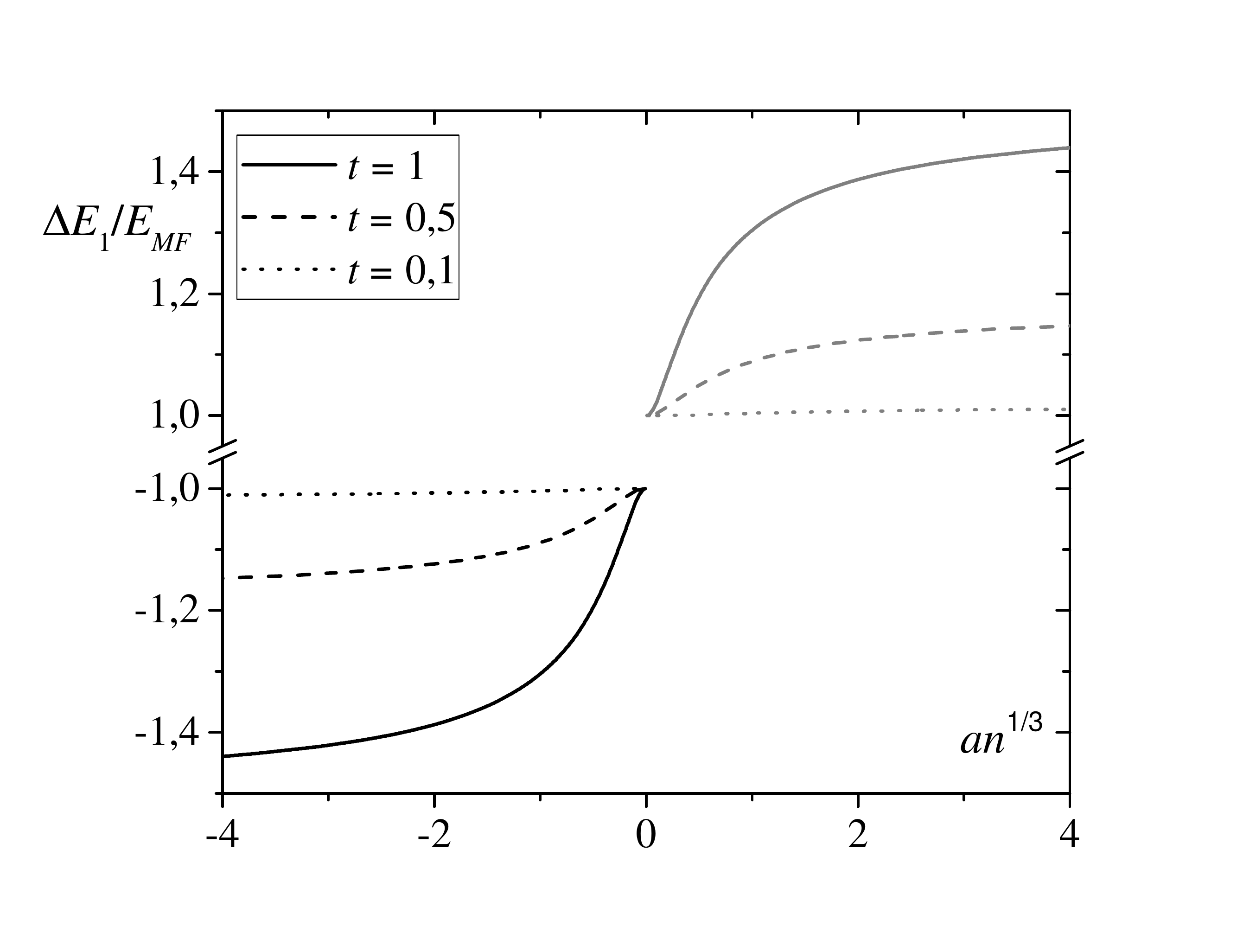}
	\caption{One-impurity energy (\ref{DeltaE1}) versus gas parameter $an^{1/3}$. Negative values of $an^{1/3}$ correspond to region, where there are no bosonic bound states and consequently the system is stable.}
\end{figure}
The leading-order temperature correction scales as $(an^{1/3})^2(T/T_0)^{5/2}$, when interaction is weak and $(T/T_0)^{3/2}$ at unitarity $|a|n^{1/3}\gg 1$.
The explicit formula for the two-impurity energy $\Delta E_2(R_{12})$ is more cumbersome (see Appendix) and therefore not written here. We can now define the effective potential energy between two impurities induced by the interaction with Bose particles as a difference of energies with fixed $R_{12}$ and infinite (one-impurity limit) distances between static particles
\begin{eqnarray}\label{Phi_eff}
\Phi^{(2)}_{\rm eff}(R_{12})=\Delta E_2(R_{12})-\Delta E_2(\infty).
\end{eqnarray}
Potential (\ref{Phi_eff}) in BEC phase has two types of terms and their origin is readily seen from the general formula for energy (\ref{E}) (see also Appendix). The first term is the temperature-independent one with a very simple mathematical expression
\begin{eqnarray}\label{Phi_eff_0}
\left.\Phi^{(2)}_{\rm eff}(R_{12})\right|_{T\to 0}=\frac{2\pi\hbar^2a}{m}n\times 2\left[\frac{1}{1+a/R_{12}}-1\right],
\end{eqnarray}
while the second term contains all thermal effects but can be calculated only numerically. Figure~4 displays the total impact of these two terms. Particularly, we have built the effective potential for only two temperatures, namely, $T=0$ and $T=T_0$, because for all other temperatures the curves describing $\Phi^{(2)}_{\rm eff}(R_{12})$ lie between those two. 
\begin{figure}[h!]
	\includegraphics[width=0.45\textwidth,clip,angle=-0]{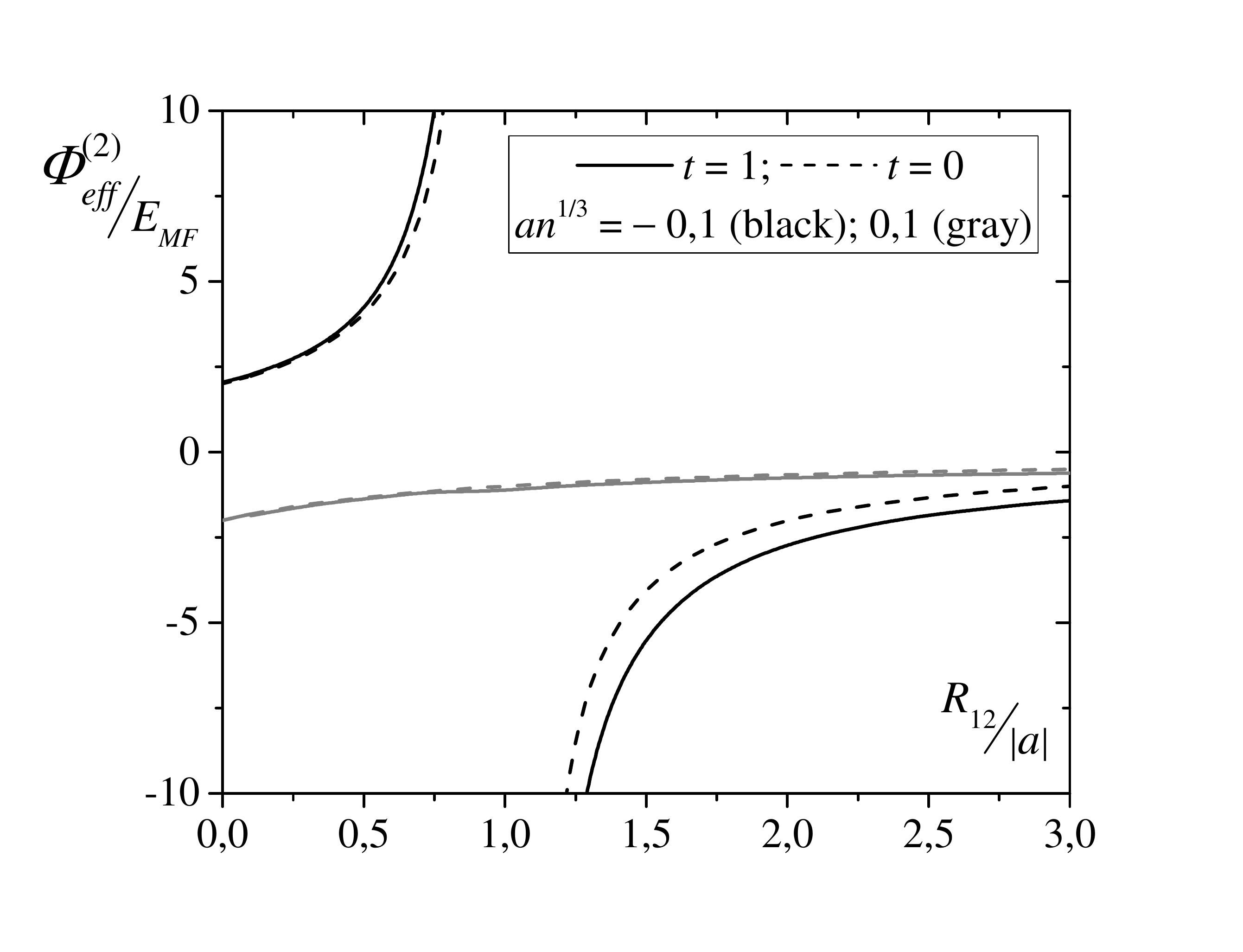}
	\includegraphics[width=0.45\textwidth,clip,angle=-0]{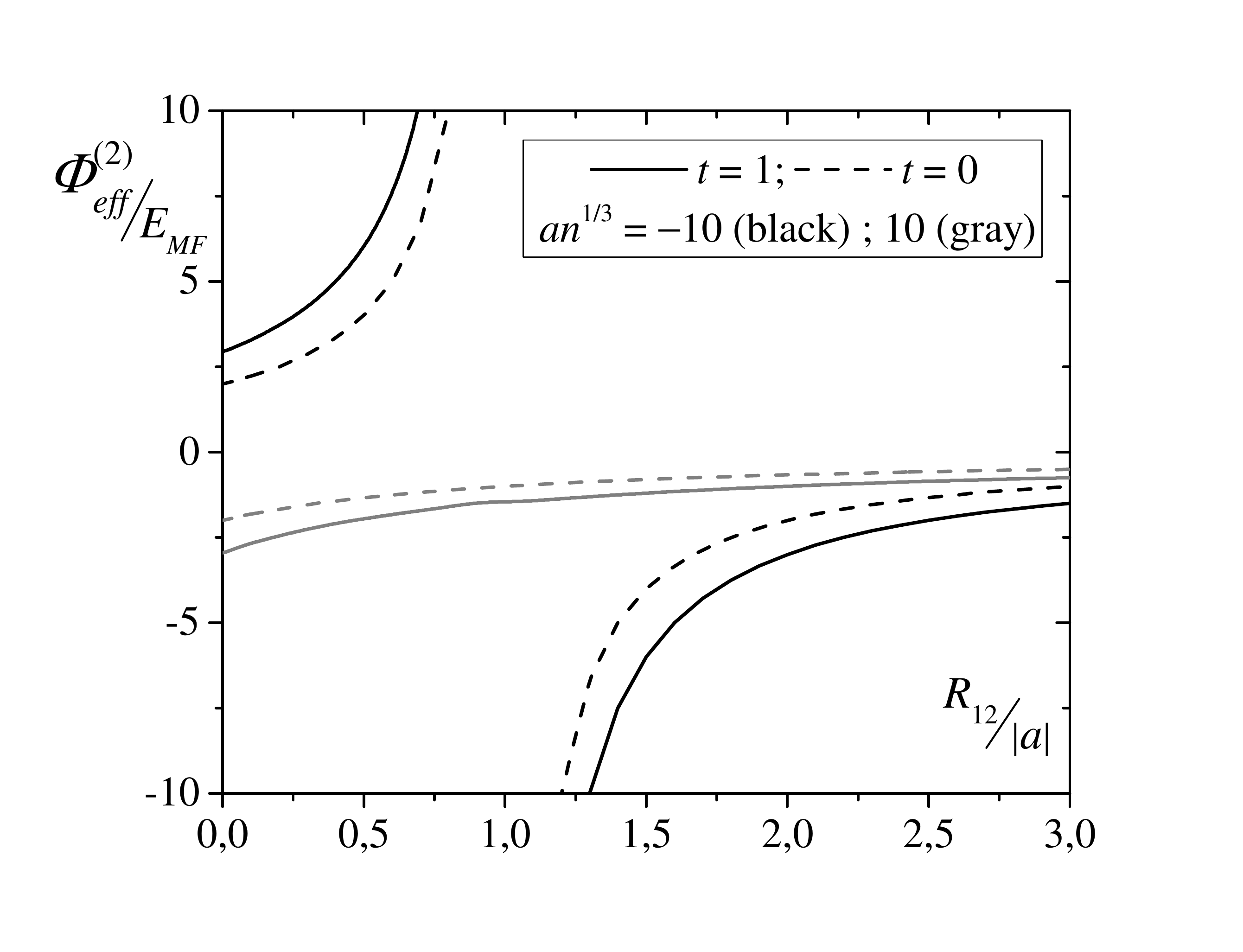}
	\includegraphics[width=0.45\textwidth,clip,angle=-0]{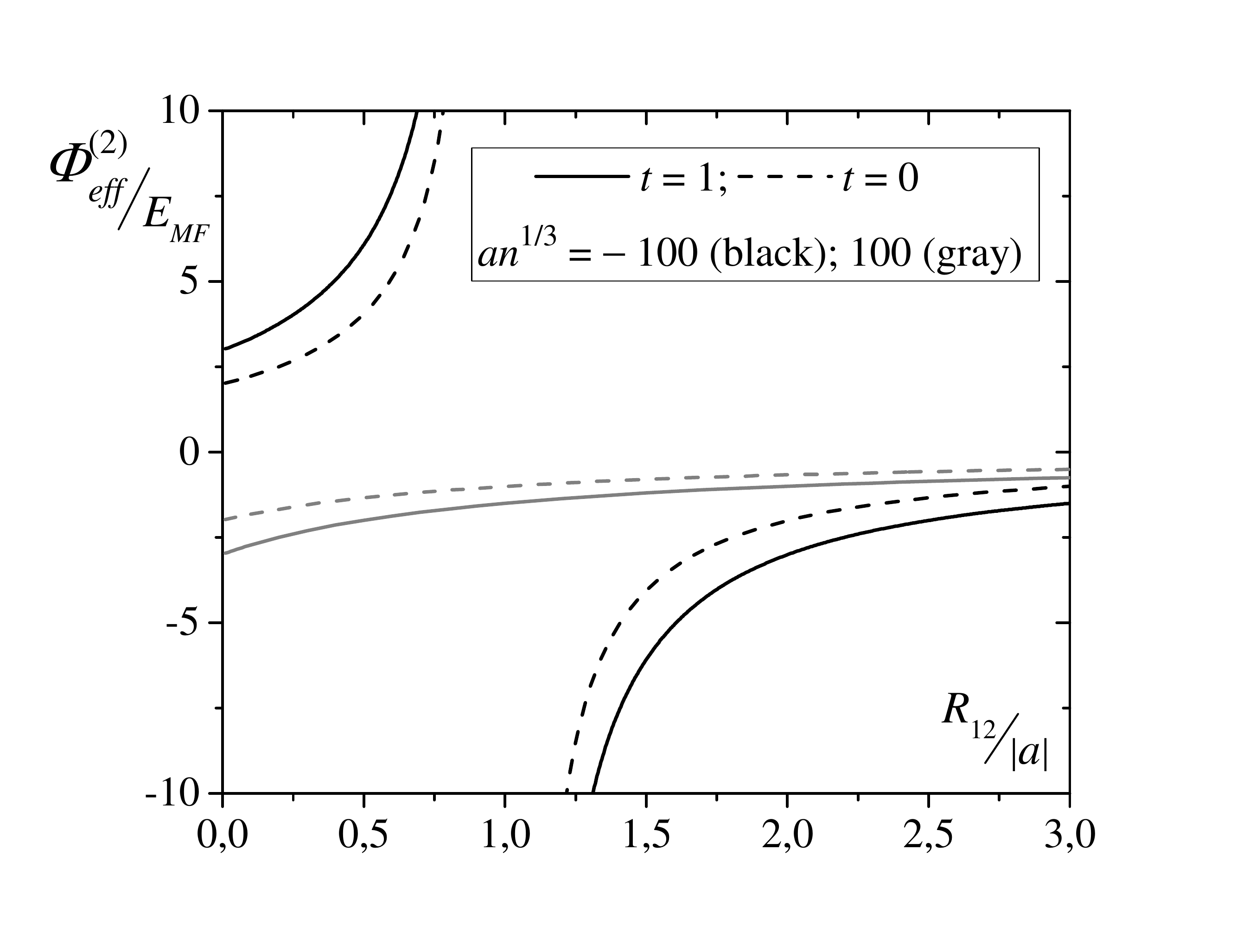}
	\caption{The effective potential $\Phi^{(2)}_{\rm eff}(R_{12})$ versus the relative distance (in units of $|a|$) between impurities at temperatures $T=0$ (dashed lines) and $T_0$ (solid lines). Different panels correspond to different boson-impurity dimensionless couplings $an^{1/3}$ from $\pm 10^{-1}$ (top) to $\pm 10^2$ (bottom) and it is clearly seen that for large $an^{1/3}$ (already from $an^{1/3}\sim 1$) fraction $\Phi^{(2)}_{\rm eff}(R_{12})/E_{{\rm MF}}$ (as function of dimensionless $R_{12}/|a|$) is insensible to changes of the interaction strength $\left.\Phi^{(2)}_{\rm eff}(R_{12})\right|_{|a|n^{1/3}\gg 1}=\left.\Phi^{(2)}_{\rm eff}(R_{12})\right|_{T\to 0}\times(1+t^{3/2}/2)$.}
\end{figure}
The presented in Fig.~4 graphs of function $\Phi^{(2)}_{\rm eff}(R_{12})$ at various interaction strengths clearly demonstrate the tendency of the Casimir forces mediated by free bosons to increase the potential well with increasing of temperature. Qualitatively, this effect can be understood by utilizing the quasiparticle picture arguments. In the simplest approximation, the effective two-body interaction between impurities that appears due to the quasiparticles exchange (regardless of sign of $a$), is more intense more bosons are in the exited states. Therefore, increase of the temperature naturally leads to increase of the quasiparticle number and, in turn, to the deepening of potential well for the attractive and raise of the potential barrier for repulsive impurity-impurity interactions.

The leading order large-distance behavior of the Casimir forces is know to be universal, i.e., controlled by critical exponents of bulk system in close vicinity of the second-order phase transition. This realizes in the classical systems, where the Casimir effect is exponentially suppressed outside the critical temperature. In Bose systems with the infinite-mass impurities immersed, in contrast, the large-$R_{12}$ asymptotic of $\Phi^{(2)}_{\rm eff}(R_{12})$ is determined by the infra-red behavior of the dynamic structure factor $S(\omega=0,k)$ in the static limit. For the case of ideal Bose gas considered here, the small-$k$ expansion $S(\omega=0,k)\propto 1/k^2$ is universal at any temperatures below $T_0$ including $T=0$. It means that universality of the Casimir forces in ideal Bose gas at low temperatures is fully dictated by presence of the Bose condensate.

The three-body effective potential principally contains three types of terms: \begin{eqnarray}\label{Phi3_eff}
&\Phi^{(3)}_{\rm eff}(R_{12},R_{23},R_{13})=\Delta E_3(R_{12},R_{23},R_{13})-3\Delta E_1\nonumber\\
&-\Phi^{(2)}_{\rm eff}(R_{12})-\Phi^{(2)}_{\rm eff}(R_{23})-\Phi^{(2)}_{\rm eff}(R_{13})
\end{eqnarray}
where from the energy $\Delta E_3(R_{12},R_{23},R_{13})$ of three impurities immersed in Bose gas one should subtract the triple energy of a single atom and the total pairwise effective interaction between impurities. An explicit expression for $\Delta E_3(R_{12},R_{23},R_{13})$ is cumbersome therefore not written. In Appendix, however, we present a formula for the energy $\Delta E_{\mathcal{N}}$ of arbitrary number $\mathcal{N}$ of impurities immersed, but more or less closed-form relation for $\Delta E_{\mathcal{N}}$ is available up to two particles. In all other cases (as that, $\mathcal{N}=3$, discussed here) the numerical procedure of calculations should be applied. At first, let us analyze simplified case of equidistant impurities (see Fig.~7). 
\begin{figure}[h!]
	\includegraphics[width=0.45\textwidth,clip,angle=-0]{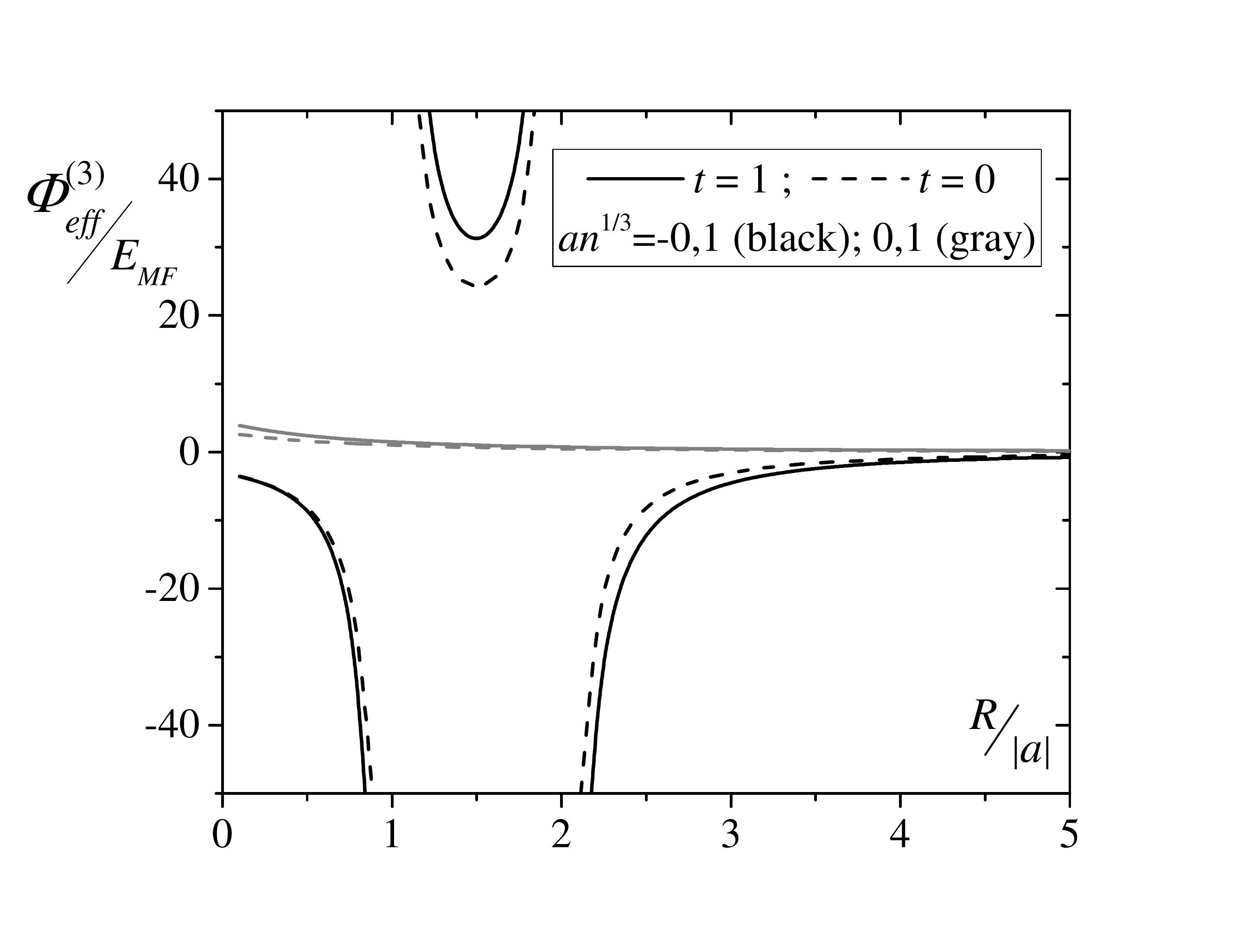}
	\includegraphics[width=0.45\textwidth,clip,angle=-0]{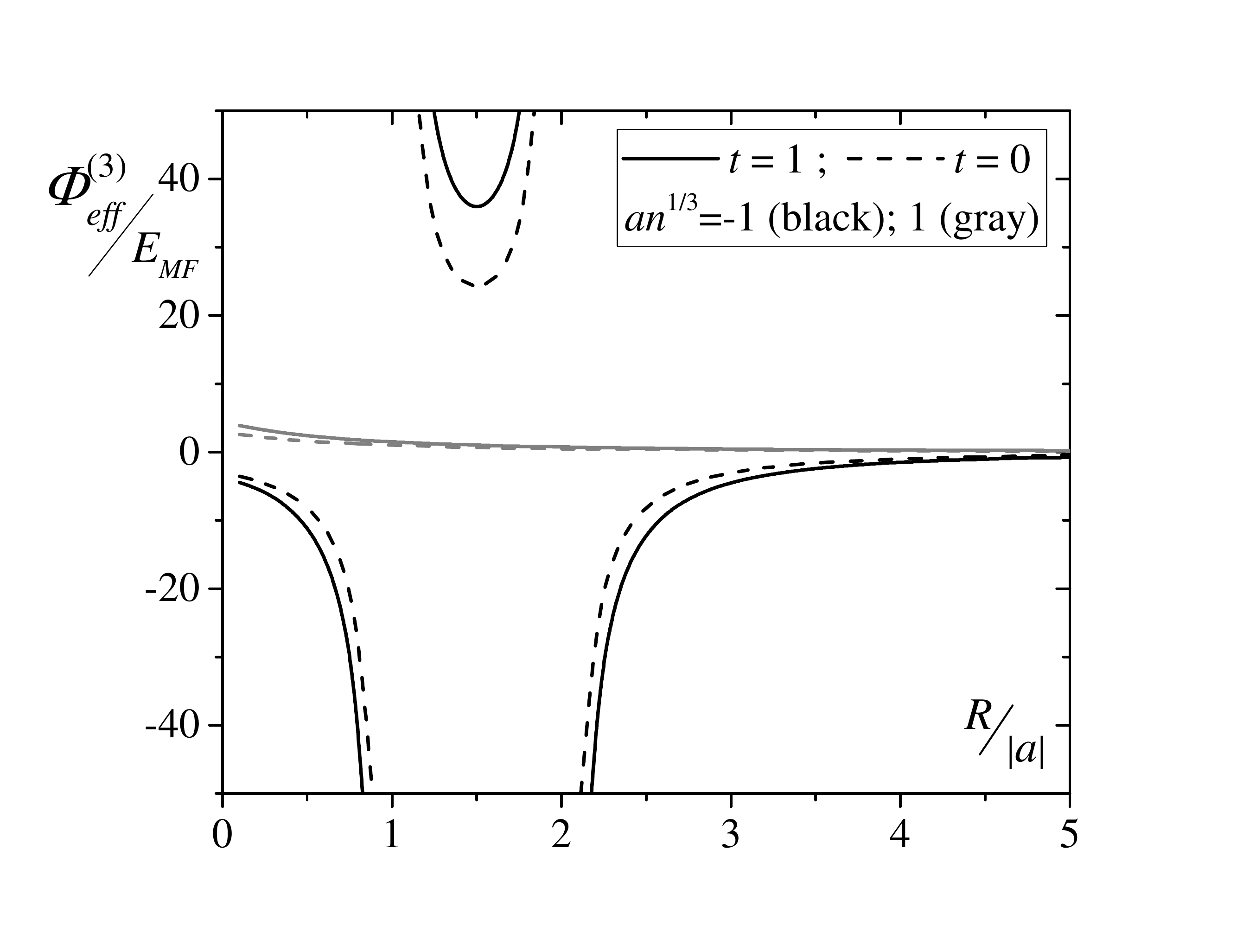}
	\includegraphics[width=0.45\textwidth,clip,angle=-0]{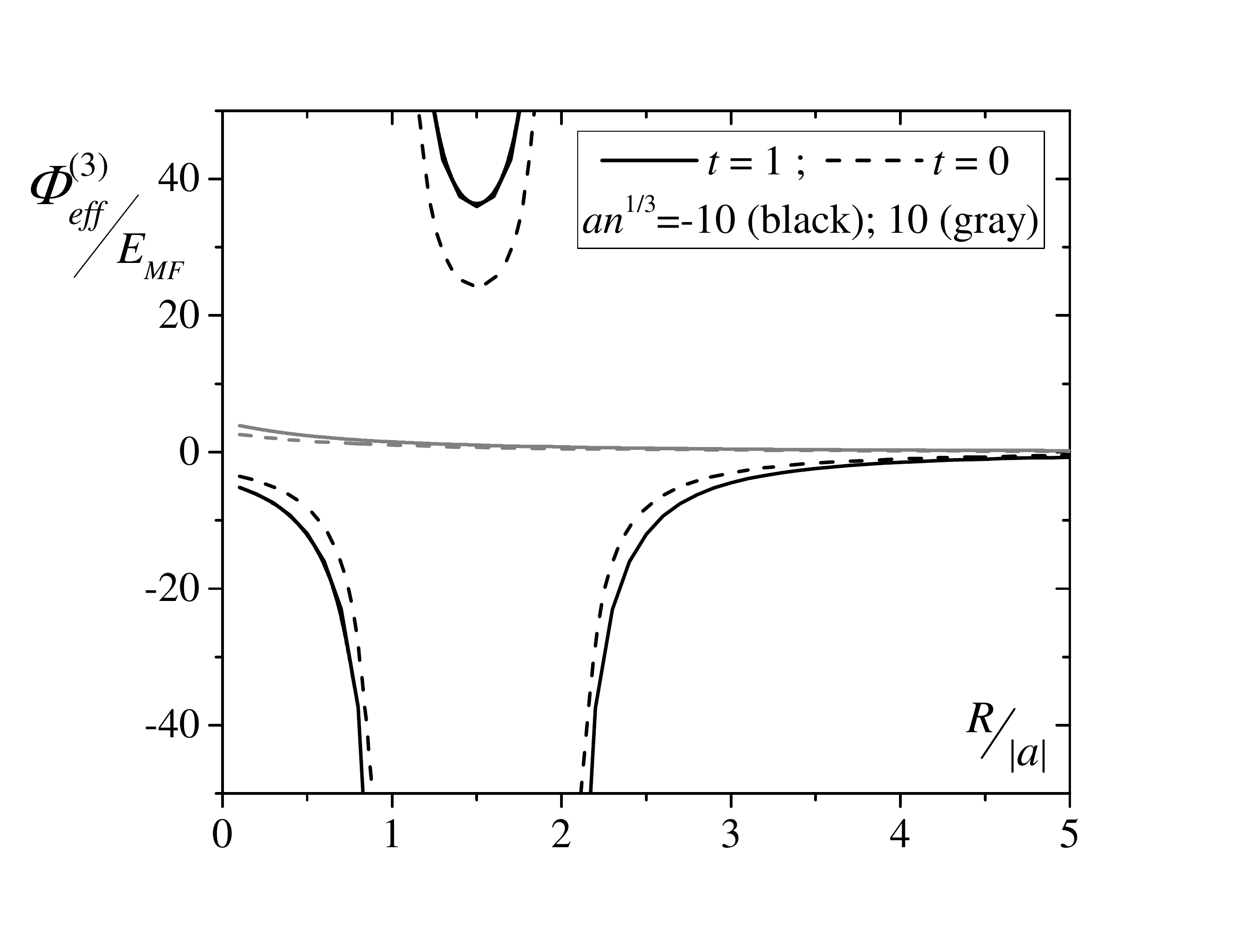}
	\caption{The effective potential $\Phi^{(3)}_{\rm eff}(R,R,R)$ versus the relative distance (in units of $|a|$) between equidistant impurities at temperatures $T=0$ (dashed lines) and $T_0$ (solid lines), respectively.}
\end{figure}
The three-body potential $\Phi^{(3)}_{\rm eff}(R)$, likewise the two-body one $\Phi^{(2)}_{\rm eff}(R_{12})$, has its own singularities that appear exclusively when $a<0$. It is intuitively clear that the three-particle potential $\Phi^{(3)}_{\rm eff}(R,R,R)$ drops to zero at large distances ($R/|a|\gg 1$) faster than $\Phi^{(2)}_{\rm eff}(R)$ both for $a<0$ and $a>0$. One also sees that the temperature effects are more decisive at $a<0$ and ratio $\left|\Phi^{(3)}_{\rm eff}(R,R,R)\right|/E_{\textrm{MF}}$ increases with the decreasing of strength of the boson-impurity interaction.

The three-body effective potential with fixed positions ($R_{12}/|a|=0.5$) of two particles is the surface which projection is plotted in Fig.~8
\begin{figure}[h!!]
	\includegraphics[width=0.5\textwidth,clip,angle=-0]{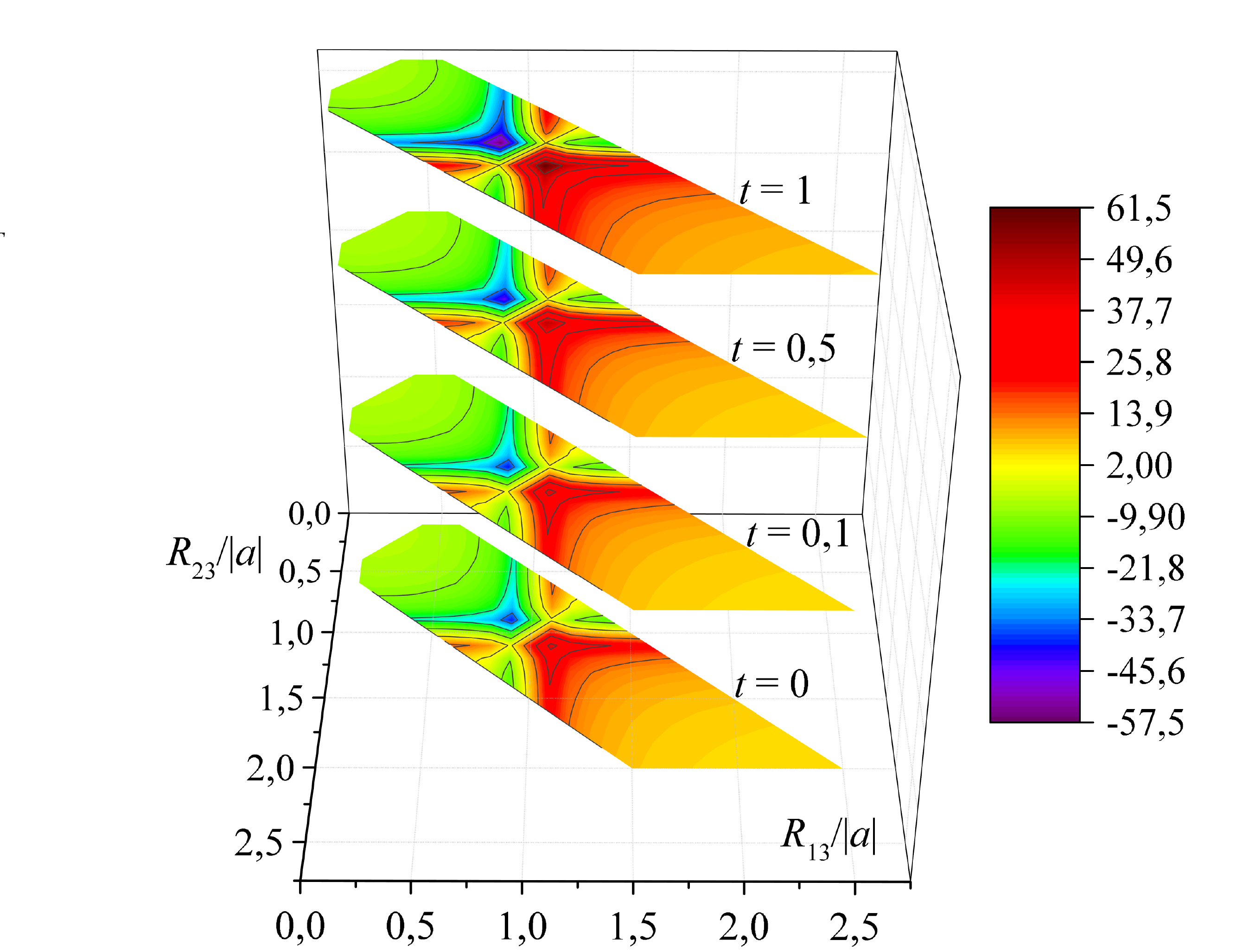}
	\includegraphics[width=0.5\textwidth,clip,angle=-0]{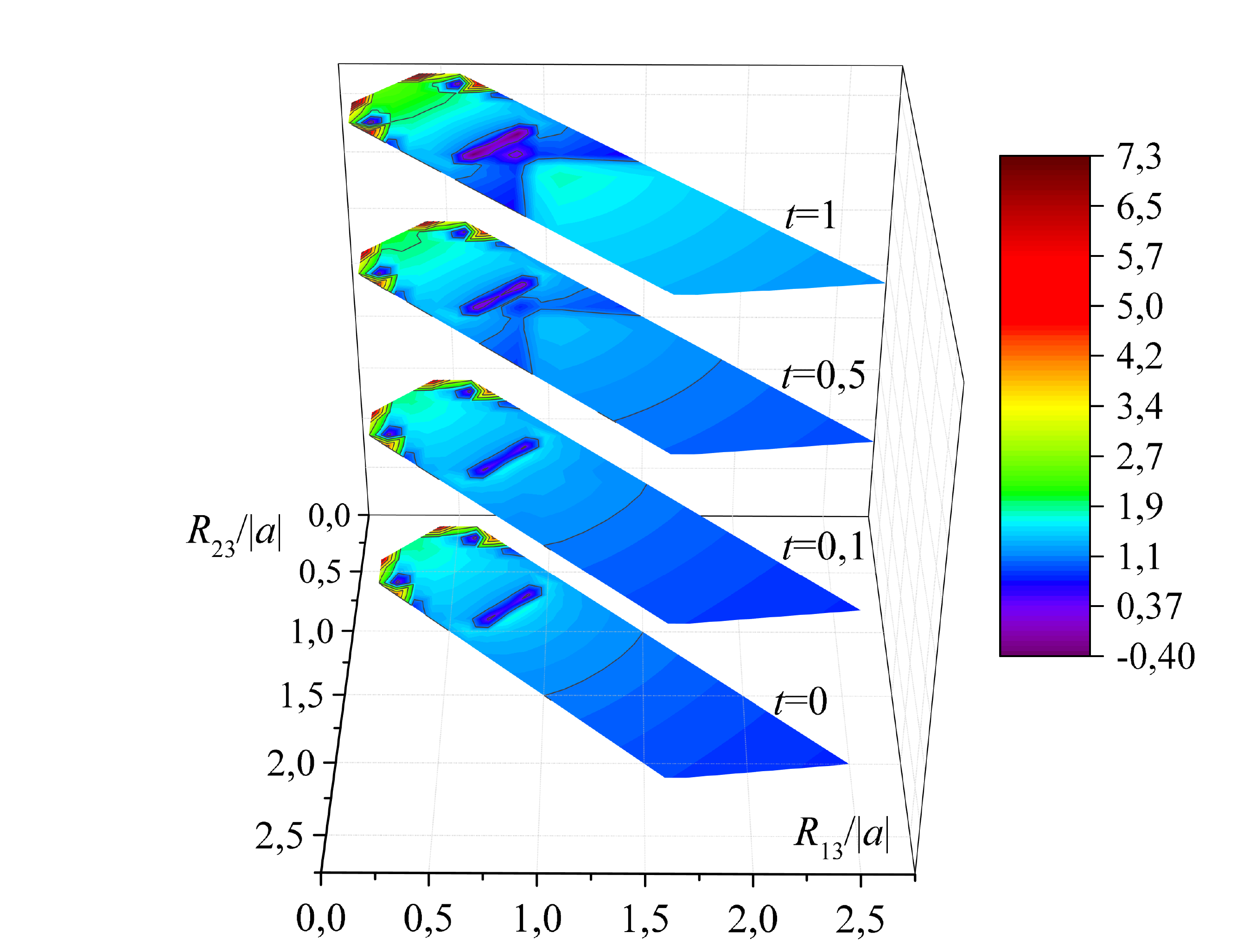}
	\caption{(Color online) The effective potential $\Phi^{(3)}_{\rm eff}(R_{12},R_{23},R_{13})$ (in units of $E_{\textrm{MF}}$).}
\end{figure}
as function of $R_{13}/|a|$ and $R_{23}/|a|$. These figures reveal two important features of the induced effective three-body potential. First, $\Phi^{(3)}_{\rm eff}(R_{12},R_{23},R_{13})$ strongly depends on a sign of the scattering length $a$. Secondly, its general behavior is readily seen at zero temperature: position of singular points ($-\infty$ at the center of violet, and $+\infty$ at the center of red regions, respectively), magnitude of the effective potential. The temperature effects, similarly to their impact on the two-body effective potential, only emphasize the general tendency of the three-body Casimir forces by increasing magnitudes of peaks and depths of wells, respectively.

\section{Conclusions}
In summary, we have calculated in detail the temperature-dependent energies associated with the immersion of one, two and three static impurities into ideal three-dimensional Bose-Einstein condensate. The simple and efficient method used here allows the exact treatment of the problem and could serve a good starting point for possible extensions on case (i) of mobile impurities and (ii) interacting Bose environments. In the former case, the exact solution exists only for 1D systems in the limit of equal masses of the immersed atom and host particles but we expect that behavior of the finite-mass impurity in 3D will be qualitatively similar to the one described here. In the present article, however, the main emphasis was made on the Casimir effect that results in the boson-mediated effective two- and three-impurity interaction and to the problem of stability of the Bose system against collapse. The latter question is very important from the point of view of preparation of such a mixture, because in contrast to non-interacting fermions, the ideal Bose gas is a substance with zero compressibility below the critical temperature. In this work we have shown that for small concentrations of uniformly-distributed impurities the system remains stable at least when the short-range boson-impurity interaction has an {\it attractive} character.

\begin{center}
	{\bf Acknowledgements}
\end{center}
We are indebted to Prof.~Andrij Rovenchak and Dr.~Artem Volosniev for many useful advices. The publication contains the results of studies conducted by President’s of Ukraine grant for competitive projects F82/205-2019.

\section{Appendix}
For completeness we give explicit analytic formulas, after integrations over the wave-vector in the thermodynamic limit, for the two-body boson-impurity $T$-matrix $t_{\omega}$ and dimensionless function $\Delta_{\omega}(R_{12})$ [see Eq.~(\ref{t}) and (\ref{Delta_R}), respectively] introduced in main text 
\begin{eqnarray*}
	t_{\omega+i0}=\frac{2\pi\hbar^2a}{m}\frac{1}{1-ak_{\omega+i0}},\\[6pt] \Delta_{\omega+i0}(R_{12})=-\frac{a}{R_{12}}\frac{e^{-R_{12}k_{\omega+i0}}}{1-ak_{\omega+i0}},
\end{eqnarray*}
where $k_{\omega+i0}=\sqrt{2m|\omega|/\hbar^2}\left\{\theta(-\omega)[1-i0]-i\theta(\omega)\right\}$ with $\theta(x)$ being the Heaviside step function.

Equation (\ref{E}) contains two types of non-vanishing terms in the thermodynamic limit, namely, the ideal Bose gas contribution which is of order $V$ and the terms of order unity corresponding to the impurities
\begin{eqnarray*}\label{E_details}
E=N_0\langle {\bf 0}|\mathcal{T}^{\bf 0}_{0}|{\bf 0}\rangle+\int^{\infty}_{\mu} \frac{{d\omega}\, \omega}{e^{(\omega-\mu)/T}-1}\\
\times\frac{-1}{\pi}\sum_{{\bf k}\neq 0}{\rm Im}\left\{\frac{1}{\omega+i0-\varepsilon_k}+\frac{\langle{\bf k}|\mathcal{T}^{{\bf 0}}_{\omega+i0}|{\bf k}\rangle}{(\omega+i0-\varepsilon_k)^2}\right\}.
\end{eqnarray*}
Shifting the integration limits, recalling that the chemical potential $\mu=\langle{\bf 0}|\mathcal{T}^{{\bf 0}}_{0}|{\bf 0}\rangle \propto 1/V$ in the BEC phase and picking up terms of order unity we obtain the energy associated with $\mathcal{N}$ impurities
\begin{eqnarray}\label{E_N}
	\Delta E_{\mathcal{N}}=N\langle {\bf 0}|\mathcal{T}^{\bf 0}_{0}|{\bf 0}\rangle+\int^{\infty}_{0}\frac{{d\omega}\,\omega}{e^{\omega/T}-1}\nonumber\\
	\times\frac{-1}{\pi}\sum_{{\bf k}\neq 0}{\rm Im}\frac{\langle{\bf k}|\mathcal{T}^{{\bf 0}}_{\omega+i0}|{\bf k}\rangle-\langle {\bf 0}|\mathcal{T}^{\bf 0}_{0}|{\bf 0}\rangle}{(\omega+i0-\varepsilon_k)^2}.
\end{eqnarray}
Integration over the wave-vector in Eq.~(\ref{E_N}) is simple, and in a case of two static particles we have
\begin{eqnarray*}
\frac{-1}{\pi}\sum_{{\bf k}\neq 0}{\rm Im}\frac{\langle{\bf k}|\mathcal{T}^{{\bf 0}}_{\omega+i0}|{\bf k}\rangle-\langle {\bf 0}|\mathcal{T}^{\bf 0}_{0}|{\bf 0}\rangle}{(\omega+i0-\varepsilon_k)^2}=-\left(\frac{m}{\pi\hbar^2}\right)^2{\rm Im}\frac{1}{k_{\omega+i0}}\\
\times\left\{t_{\omega+i0}\frac{1+\Delta_{\omega+i0}(R_{12})e^{-R_{12}k_{\omega+i0}}}
{1-\Delta^2_{\omega+i0}(R_{12})}-\frac{t_0}{1-\Delta_0(R_{12})}\right\}.
\end{eqnarray*}
Taking the imaginary part, substituting in Eq.~(\ref{E_N}) and making use of change of variables in the integral, we arrive with energy $\Delta E_2(R_{12})$ as a function of temperature and relative distance $R_{12}$ between particles
\begin{eqnarray}
	&&\Delta E_{2}(R_{12})=\frac{4\pi\hbar^2an}{m}\frac{1}{1+a/R_{12}}\nonumber\\
	&&-\frac{4\pi\hbar^2a}{m}\int_0^\infty\frac{dk}{(2\pi)^2}\frac{k^2}{e^{\varepsilon_k/T}-1}\nonumber\\
	&&\times\Bigg[
	\frac{1+a\cos(kR_{12})/R_{12}}{(1+a\cos(kR_{12})/R_{12})^2+(ak+a\sin(kR_{12})/R_{12})^2}\nonumber\\
	&&+\frac{1}{2}\frac{ak\sin(kR_{12})-(1-a/R_{12})(1-\cos(kR_{12}))}{(1+a\cos(kR_{12})/R_{12})^2+(ak+a\sin(kR_{12})/R_{12})^2}\nonumber\\
	&&-\frac{1}{2}\frac{ak\sin(kR_{12})-(1+a/R_{12})(1-\cos(kR_{12}))}{(1-a\cos(kR_{12})/R_{12})^2+(ak-a\sin(kR_{12})/R_{12})^2}\nonumber\\
	&&-\frac{1}{1+a/R_{12}}
	\Bigg].
\end{eqnarray}
Putting $R_{12}\to \infty$ we recover the doubled binding energy of a single impurity $\Delta E_2(\infty)=2\Delta E_1$, with $\Delta E_1$ presented in main text.


\begin{thebibliography}{99}
	
\bibitem{Catani} J.~Catani, G.~Lamporesi, D.~Naik, M.~Gring, M.~Inguscio, F.~Minardi, A.~Kantian, and T.~Giamarchi, 
\href{https://doi.org/10.1103/PhysRevA.85.023623}{Phys.~Rev.~A {\bf 85}, 023623 (2012).}
\bibitem{Jorgensen} N.~B.~Jorgensen, L.~Wacker, K.~T.~Skalmstang, M.~M.~Parish, J.~Levinsen, R.~S.~Christensen, G.~M.~Bruun and J.~J.~Arlt,
\href{https://journals.aps.org/prl/abstract/10.1103/PhysRevLett.117.055302}{Phys.~Rev.~Lett.  {\bf 117}, 055302 (2016).}

\bibitem{Hu} M.-G.~Hu, M.~J.~Van de Graaff, D.~Kedar, J.~P.~Corson, E.~A.~Cornell and D.~S.~Jin, 
\href{https://journals.aps.org/prl/abstract/10.1103/PhysRevLett.117.055301}{Phys.~Rev.~Lett. {\bf 117}, 055301 (2016).}
\bibitem{Yan} Z.~Z.~Yan, Y.~Ni, C.~Robens, M.~W.~Zwielein, 
\href{https://science.sciencemag.org/content/368/6487/190/tab-article-info}{Science  
	{\bf  368}, 190 (2020)}
\bibitem{Camargo} F.~Camargo, R.~Schmidt, J.~D.~Whalen, R.~Ding, G.~Woehl Jr., S.~Yoshida, J.~Burgdorfer, F.~B.~Dunning, H.~R.~Sadeghpour, E.~Demler and T.~C.~Killian, 
\href{https://journals.aps.org/prl/abstract/10.1103/PhysRevLett.120.083401}{Phys.~Rev.~Lett. {\bf 120}, 088401 (2018).}


\bibitem{Schmid} S.~Schmid, A.~Harter, J.~H.~Denschlag,  \href{https://doi.org/10.1103/PhysRevLett.105.133202}{Phys.~Rev.~Lett. {\bf 105}, 133202 (2010).}

\bibitem{Levinsen1} J.~Levinsen, M.~M.~Parish, R.~S.~Christensen, J.~J.~Arlt, and G.~M.~Bruun, 
\href{https://doi.org/10.1103/PhysRevA.96.063622}{Phys.~Rev.~A. {\bf 96}, 063622 (2017).}

\bibitem{Guenther} N.~E.~Guenther, P.~Massignan, M.~Lewenstein, and G.~M.~Bruun, 
\href{https://doi.org/10.1103/PhysRevLett.120.050405}{Phys.~Rev.~Lett. {\bf 120}, 050405 (2018).}

\bibitem{Pastukhov2018} V.~Pastukhov, 
\href{https://doi.org/10.1088/1751-8121/aab9c1}
{J.~Phys.~A: Math.~Theor. {\bf 51}, 195003 (2018).}

\bibitem{Boudjemaa} A.~Boudjemaa,  \href{https://doi.org/10.1088/1751-8113/48/4/045002}{J. Phys. A: Math. Theor. {\bf 48}, 045002 (2015).}

\bibitem{Liu} W.~E.~Liu, J.~Levinsen, and M.~M.~Parish
\href{https://doi.org/10.1103/PhysRevLett.122.205301}{Phys. Rev. Lett. {\bf 122}, 205301 (2019).}

\bibitem{Mehboudi} M.~Mehboudi, A.~Lampo, C.~Charalambous, L.~A.~Correa, M.~A.~Garcia-March, and M.~Lewenstein, 
\href{https://doi.org/10.1103/PhysRevLett.122.030403}{Phys. Rev. Lett. {\bf 122}, 030403 (2019).}

\bibitem{Naidon} P.~Naidon,  \href{https://doi.org/10.7566/JPSJ.87.043002}{J.~Phys.~Soc.~Jpn. {\bf 87} 043002 (2018).}

\bibitem{Tempere} W.~Casteels, J.~Tempere, J.~T.~Devreese, 
\href{https://doi.org/10.1103/PhysRevA.88.013613}{Phys.~Rev.~A {\bf 88} 013613 (2013).}

\bibitem{Ardila} A.~Camacho-Guardian, L.~A.~Pena Ardila, T.~Pohl, G.~M.~Bruun, \href{https://doi.org/10.1103/PhysRevLett.121.013401}{Phys.~Rev.~Lett. {\bf 121} 013401 (2018).}


\bibitem{Levinsen2} J.~Levinsen, M.~M.~Parish, and G.~M.~Bruun, 
\href{https://doi.org/10.1103/PhysRevLett.115.125302}
{Phys. Rev. Lett. {\bf 115}, 125302 (2015).}

\bibitem{Sun} M.~Sun, H.~Zhai, and X.~Cui,
\href{https://doi.org/10.1103/PhysRevLett.119.013401}
{ Phys. Rev. Lett. {\bf 119}, 013401 (2017).}

\bibitem{Zinner2013} N.~T.~Zinner,  \href{https://doi.org/10.1209/0295-5075/101/60009}{Euro.~Phys.~Lett. {\bf 101}, 60009 (2013).}
\bibitem{Zinner2014} N.~T.~Zinner, 
\href{https://doi.org/10.1140/epjd/e2014-40842-y}{Euro.~Phys.~J.~D {\bf 68}, 216 (2014).}


\bibitem{Klimchitskaya} G.~L.~Klimchitskaya, U.~Mohideen, V.~M.~Mostapenko,  \href{https://doi.org/10.1103/RevModPhys.81.1827}{Rev.~Mod.~Phys. {\bf 81}, 1827 (2009).}

\bibitem{Recati} A.~Recati, J.~N.~Fuchs, C.~S.~Peca, W.~Zwerger, \href{https://doi.org/10.1103/PhysRevA.72.023616}{Phys.~Rev.~A {\bf 72}, 023616 (2005).}

\bibitem{Wachter} P.~Wachter, V.~Meden, K.~Schonhamer, 
\href{https://doi.org/10.1103/PhysRevB.76.045123}{Phys.~Rev.~B {\bf 76}, 045123 (2007).}

\bibitem{Kamenev} M.~Schecter, A.~Kamenev, 
\href{https://doi.org/10.1103/PhysRevLett.112.155301}{Phys.~Rev.~Lett. {\bf 112}, 155301 (2014).}

\bibitem{Mistakidis2019} S.~I.~Mistakidis, L.~Hilbig, and P.~Schmelcher,  
\href{https://doi.org/10.1103/PhysRevA.100.023620}
{Phys. Rev. A {\bf 100}, 023620 (2019).}

\bibitem{Rodin} A. Rodin, 
\href{https://journals.aps.org/prb/abstract/10.1103/PhysRevB.100.195403}{Phys. Rev. B {\bf 100}, 195403 (2019)}


\bibitem{Pavlov2018} A.~I.~Pavlov, J.~Brink, D.~V.~Efremov,  
\href{https://doi.org/10.1103/PhysRevB.98.161410}
{Phys. Rev. B {\bf 98}, 161410(R) (2018).}

\bibitem{Pavlov2019}A.~I.~Pavlov, J.~Brink, D.~V.~Efremov,  
\href{https://doi.org/10.1103/PhysRevB.100.014205}
{Phys. Rev. B {\bf 100}, 014205 (2019).}


\bibitem{Salvo} B.~J.~DeSalvo, K.~Patel, G.~Gai, C.~Chin,  
\href{https://doi.org/10.1038/s41586-019-1055-0}
{Nature {\bf 568}, 61 (2019).}

\bibitem{Nishida} Y.~Nishida, 
\href{https://doi.org/10.1103/PhysRevA.79.013629}
{Phys.~Rev.~A {\bf 79}, 013629 (2009).}

\bibitem{Song} Pei-Song He, Qing Sun, and An-Chun Ji, 
\href{https://doi.org/10.1103/PhysRevA.96.043617}
{Phys. Rev. A {\bf 96}, 043617 (2017).}

\bibitem{Guardian} A.~Camacho-Guardian and G.~M.Bruun,  \href{https://doi.org/10.1103/PhysRevX.8.031042}
{Phys.~Rev.~X. {\bf 8}, 031042 (2018).}


\bibitem{Dehkharghani} A.~S.~Dehkharghani, A.~G.~Volosniev, and N.~T.~Zinner,
\href{https://doi.org/10.1103/PhysRevLett.121.080405}{ Phys.~Rev.~Lett. {\bf 121}, 080405 (2018).}

\bibitem{Reichert1} B.~Reichert, Z.~Ristivojevic, A.~Petkovic, 
\href{https://doi.org/10.1088/1367-2630/ab1b8e}
{New.~J.~Phys. {\bf 21}, 053024 (2019).}

\bibitem{Reichert3} B.~Reichert, Z.~Ristivojevic, A.~Petkovic,
\href{https://journals.aps.org/prb/abstract/10.1103/PhysRevB.100.235431}{ Phys. Rev. B {\bf 100}, 235431 (2019)}

\bibitem{Reichert2} B.~Reichert, Z.~Ristivojevic, A.~Petkovic, 
\href{https://doi.org/10.1103/PhysRevB.99.205414}{Phys. Rev. B {\bf 99}, 205414 (2019).}

\bibitem{Landau_5} L.~D.~Landau, E.~M.~Lifshitz, {\it Statistical Physics: Vol. 5} (Elsevier, 2013)

\bibitem{Luscher} M.~L{\"u}scher,  
\href{https://projecteuclid.org/euclid.cmp/1104115329}
{Commun. Math. Phys. {\bf 105}, 153 (1986).}


\end{thebibliography}
\end{document}